\documentclass[12pt]{article}
\usepackage{setspace}
\usepackage{amsmath,amsthm,amssymb,amsfonts,cancel}
\usepackage{graphicx}
\usepackage{wrapfig}
\usepackage{algorithm} 
\usepackage{algpseudocode}
\usepackage[font={small}]{caption}
\usepackage{subcaption}
\usepackage{sidecap}
\usepackage{epstopdf}
\usepackage{color}
\usepackage[usenames,dvipsnames,svgnames,table]{xcolor}
\usepackage{hyperref}
\hypersetup{hidelinks}
\usepackage{algcompatible}
\usepackage{setspace}
\onehalfspace

\usepackage{natbib}

\title{A Joint Graph Inference Case Study: \\the \textit{C.elegans} Chemical and Electrical Connectomes}

\author{Li Chen\thanks{Department of Applied Mathematics and Statistics, Johns Hopkins University, Baltimore, MD 21218, USA
      corresponding author: lchen87@jhu.edu} \and Joshua T. Vogelstein\thanks{Department of Biomedical Engineering and Institute for Computational Medicine, Johns Hopkins University, Baltimore, MD 21218, USA} \and Vince Lyzinski\thanks{Johns Hopkins University Human Language Technology Center of Excellence, Baltimore, MD, 21218, USA} \and Carey E. Priebe\thanks{Department of Applied Mathematics and Statistics, Johns Hopkins University, Baltimore, MD 21218, USA}}

\date{\today}

\begin{document}
\maketitle

\begin{abstract}
We investigate joint graph inference for the chemical and electrical connectomes of the \textit{Caenorhabditis elegans} roundworm. The \textit{C.elegans} connectomes consist of $253$ non-isolated neurons with known functional attributes, and there are two types of synaptic connectomes, resulting in a pair of graphs. We formulate our joint graph inference from the perspectives of seeded graph matching and joint vertex classification. Our results suggest that connectomic inference should proceed in the joint space of the two connectomes, which has significant neuroscientific implications.
\end{abstract}

The \textit{Caenorhabditis elegans} (\textit{C.elegans}) is a non-parasitic, transparent roundworm approximately one millimeter in length. The majority of \textit{C.elegans} are female hermaphrodites. \cite{maupas1901modes} first described the worm in 1900 and named it \textit{Rhabditis elegans}. It was later categorized under the subgenus \textit{Caenorhabditis} by \cite{osche1952bedeutung}, and then, in 1955, raised to the generic status by Ellsworth Dougherty, to whom much of the recognition for choosing \textit{C.elegans} as a model system in genetics is attributed \citep{riddle1997origins}. The long name of this nematode mixes Greek and Latin, where \textit{Caeno} means recent, \textit{rhabditis} means rod-like, and \textit{elegans} means elegant.

Research on \textit{C.elegans} rose to prominence after the nematode was adopted as a model organism: an easy-to-maintain non-human species widely studied,  so that discoveries on this model organism might offer insights for the functionality of other organisms. The discoveries of caspases \citep{yuan1993c}, RNA interference \citep{fire1998potent}, and microRNAs \citep{lee1993c} are among some of the notable research using \textit{C.elegans}.

Connectomes, the mapping of neural connections within the nervous system of an organism, provide a comprehensive structural characterization of the neural network architecture, and represent an essential foundation for basic neurobiological research. Applications based on the discovery of the connectome patterns and the identification of neurons based on their connectivity structure give rise to significant challenges and promise important impact on neurobiology. Recently there has been an increasing interest in the network properties of \textit{C.elegans} connectomes. The hermaphrodite \textit{C.elegans} is the only organism with a fully constructed connectome \citep{sulston1983embryonic}, and has one of the most highly studied nervous systems.
  
Studies on the \textit{C.elegans} connectomes traditionally focus on utilizing one single connectome alone \citep{varshney2011structural, pavlovic2014stochastic, sulston1983embryonic, towlson2013rich}, although there are many connectomes available. Notably, \citet{varshney2011structural} discovered structural properties of the \textit{C.elegans} connectomes via analyzing the connectomes' graph statistics. \citet{pavlovic2014stochastic} estimated the community structure of the connectomes, and their findings are compatible with known biological information on the \textit{C.elegans} nervous system. 

Our new statistical approach of joint graph inference looks instead at jointly utilizing the paired chemical and electrical connectomes of the hermaphrodite \textit{C.elegans}. We formulate our inference framework from the perspectives of seeded graph matching and joint vertex classification, which we will explain in Section 2. This framework gives a way to examine the structural similarity preserved across multiple connectomes within species, and make quantitative comparisons between joint connectome analysis and single connectome analysis. We found that the optimal inference for the information-processing properties of the connectome should proceed in the joint space of the \textit{C.elegans} connectomes, and using the joint connectomes predicts neuron attributes more accurately than using either connectome alone.
\section{The Hermaphrodite \textit{C.elegans} Connectomes}

The hermaphrodite \textit{C.elegans} connectomes consist of $302$ labeled neurons for each organism. The \textit{C.elegans} somatic nervous system has 279 neurons connecting to each other across synapses. There are many possible classifications of synaptic types. Here we consider two types of synaptic connections among these neurons: chemical synapses and electrical junction potentials. These two types of connectivity result in two synaptic connectomes consisting of the same set of neurons. 

We represent the connectomes as graphs. A graph is a representation of a collection of interacting objects. The objects are referred to as nodes or vertices. The interactions are referred to edges or links. In a connectome, the vertices represent neurons, and the edges represent synapses. Mathematically, a graph $G = (V , E)$ consists of a set of vertices or nodes $V = [n] := \{1, 2, ..., n\}$ and a set of edges $E \subset$. If the graph is undirected and non-loopy, then the edge set  $E= {V \choose 2} = {[n] \choose 2}$, where $\binom{[n]}{2}$ denotes all (unordered) pairs $\{v_i, v_j\}$ for all $v_i, v_j \in V$ and $v_i \neq v_j$. If the graph is directed and non-loopy, there are $n(n-1)$ possible edges, and we can represent them as ordered pairs. In this work, we assume the graphs are undirected, weighted, and non-loopy.
The adjacency matrix $A$ of $G$ is the $n$-by-$n$ matrix in which each entry $A_{ij}$ denotes the edge existence between vertex $i$ and $j$: $A_{ij}=1$ if an edge is present, and $A_{ij}=0$ if an edge is absent. The adjacency matrix is symmetric, binary and hollow, i.e., the diagonals are all zeros. 

For the hermaphrodite \textit{C.elegans} worm, the chemical connectome $G_c$ is weighted and directed. The electrical gap junctional connectome $G_g$ is weighted and undirected. This is consistent with an important characteristic of electrical synapses -- they are bidirectional \citep{purves2001neuroscience}. The chemical connectome $G_c$ has $3$ loops and no isolated vertices, while the electrical gap junctional connectome $G_g$ has no loops and 26 isolated vertices. Both connectomes are sparse. The chemical connectome $G_c$ has 2194
directed edges out of $279\cdot 278$ possible ordered neuron pairs, resulting in a sparsity level of approximately $2.8\%$. The electrical gap junctional connecome $G_g$ has 514 undirected edges out of ${279\choose 2}$ possible unordered neuron pairs, resulting in a sparsity level of approximately $1.3\%$. 

In our analysis, we are interested in the $279-26 = 253$ non-isolated neurons in the hermaphrodite \textit{C.elegans} somatic nervous system. Each of these $253$ neurons can be classified in a number of ways, including into 3 non-overlapping connectivity based types:
sensory neurons ($27.96\%$), interneurons ($29.75\%$) and motor neurons ($42.29\%$). 
Here we will work with binary, symmetric and hollow adjacency matrices of the neural connectomes throughout.
We symmetrize $A$ by $A \leftarrow A + A^T$, then binarize $A$ by thresholding the positive entries of $A$ to be 1 and 0 otherwise, and finally set the diagonal entries of $A$ to be zero. Indeed, we focus on the existence of synaptic connectomes, and the occurrence of loops is low (3 loops in $G_c$ and none in $G_g$) so we can ignore it.

\begin{figure}[h]
\centering
\includegraphics[width=0.55\textwidth]{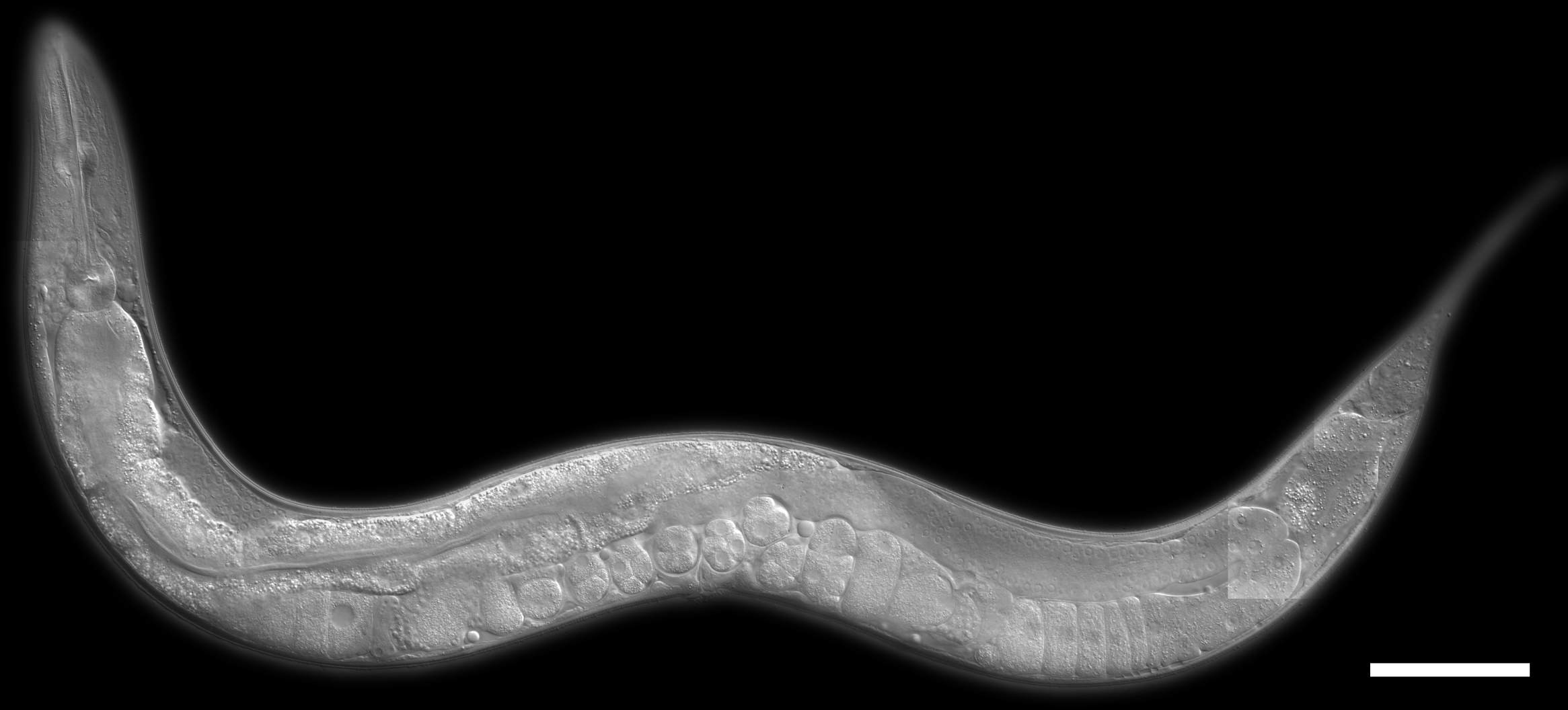}
\caption{\textbf{An image of the \textit{Caenorhabditis elegans} (\textit{C.elegans}) roundworm.} The image is available at \url{http://post.queensu.ca/~chinsang/research/c-elegans.html}.}
\label{fig: worm_body}
\end{figure}

\begin{figure}[h]
\centering
\includegraphics[scale = .17]{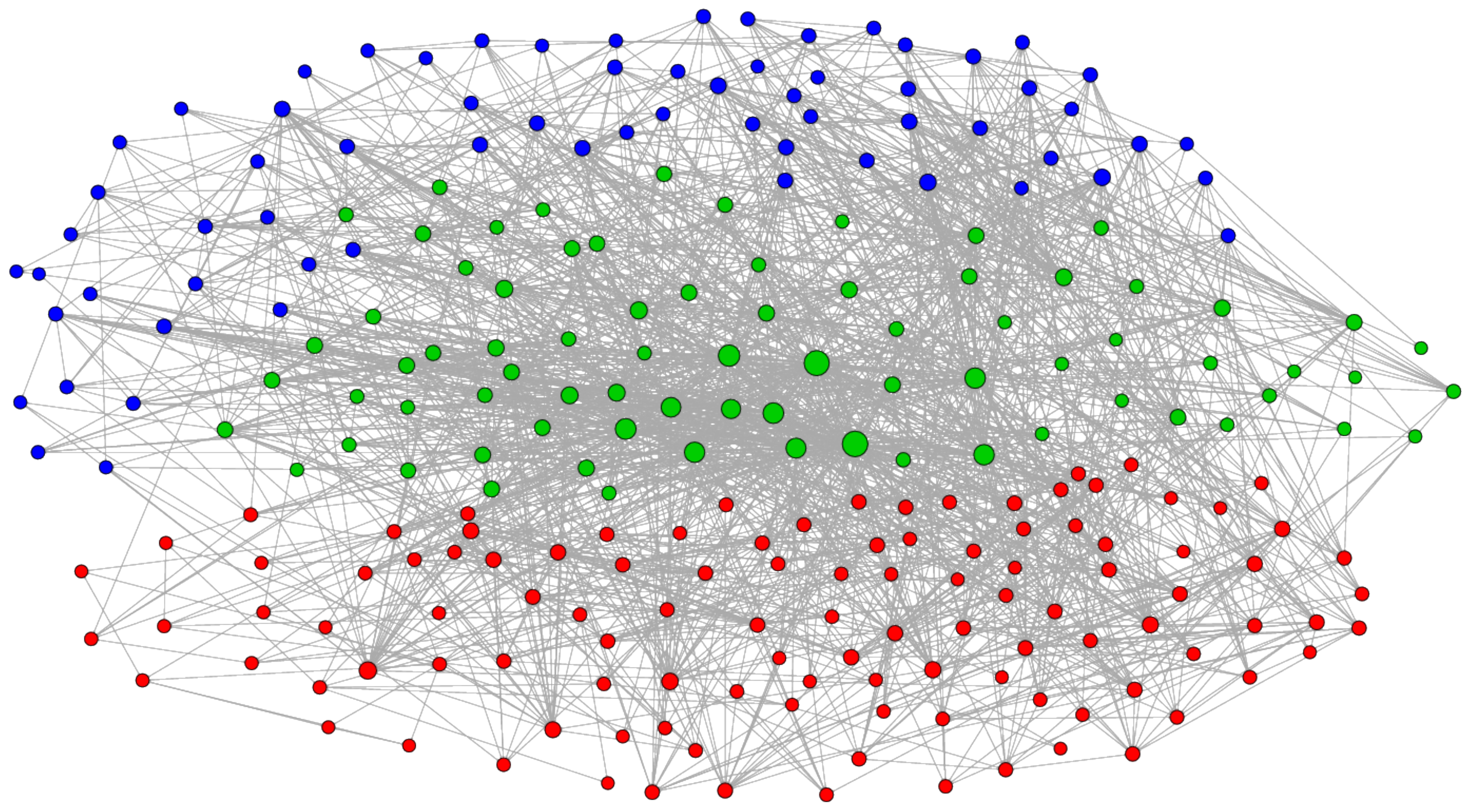}
\includegraphics[scale = .17]{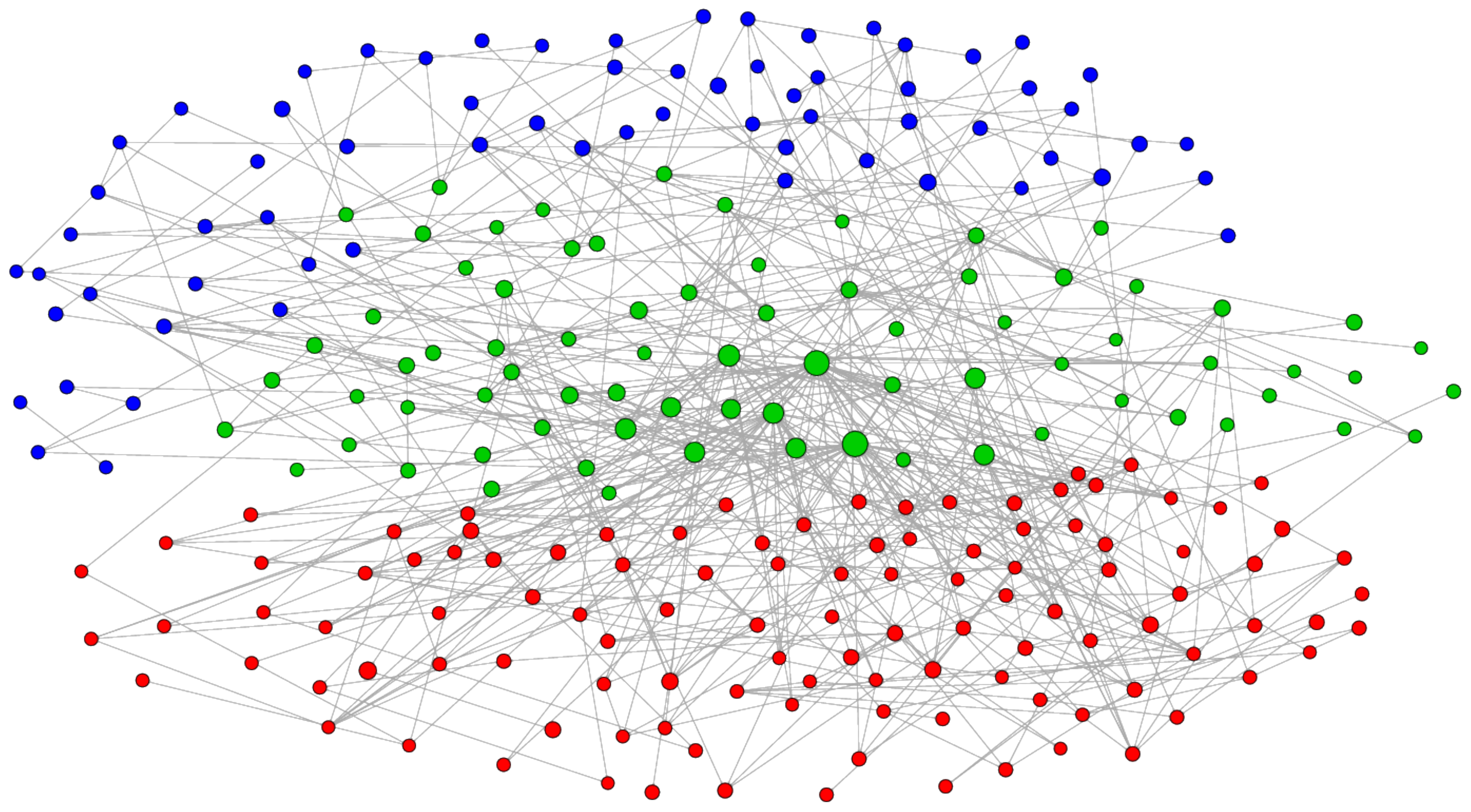} \\

\caption{\textbf{The pair of \textit{C.elegans} neural connectomes visualized as graphs.} Red nodes correspond to motor neurons, green nodes correspond to interneurons, and blue nodes correspond to sensory neurons. (Left) The chemical connectome $G_{\text{c}}$. 
(Right) The electrical gap junctional connectome $G_{\text{g}}$. 
Both synaptic connectomes are sparse, while $G_{\text{g}}$ is much sparser than $G_{\text{c}}$. A similar connectivity pattern is seen across both connectomes.}
\label{fig:pair_visual}
\end{figure}

\begin{figure}[H]
\centering
\includegraphics[scale=.4]{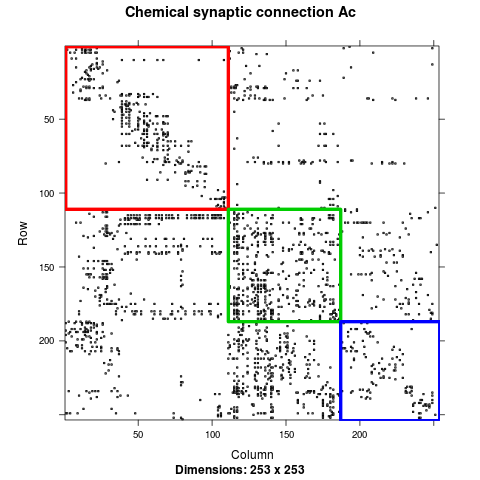} \hspace{1in}
\includegraphics[scale=.4]{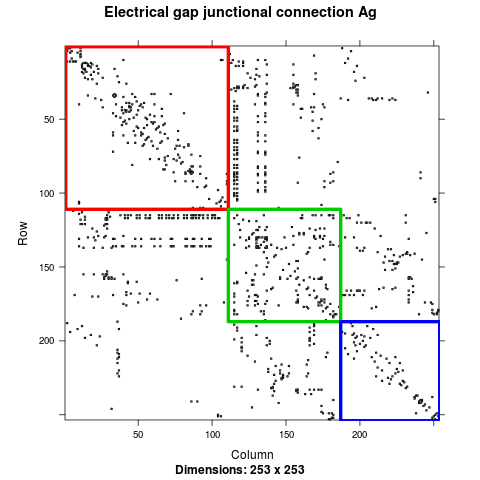}\\
\caption{(Left): The adjacency matrix $A_\text{c}$ of $G_{\text{c}}$ sorted according to the neuron types. (Right): The adjacency matrix $A_\text{g}$ of $G_{\text{g}}$ sorted according to the neuron types. The red block corresponds to the connectivity among the motor neurons, the green block corresponds to the connectivity among the interneurons, and the blue block corresponds to the connectivity among the sensory neurons.}
\label{fig:pair_adj}
\end{figure}

An image of the \textit{C.elegans} worm body is seen in Figure \ref{fig: worm_body}. The pair of the neural connectomes are visualized in Figure \ref{fig:pair_visual}. In the chemical connectome $G_{\text{c}}$, the interneurons are heavily connected to the sensory neurons. The sensory neurons are connected more frequently to the motor neurons and interneurons than amongst themselves. In the electrical gap junction potential connectome $G_{\text{g}}$, the motor neurons are heavily connected to the interneurons. The sensory neurons are connected more frequently to the motor neurons and interneurons than among themselves. The connectome dataset is accessible at \url{http://openconnecto.me/herm-c-elegans}. Figure \ref{fig:pair_adj} presents the adjacency matrices of the paired \textit{C.elegans} connectomes.

\section{Joint Graph Inference}

We consider an inference framework in the joint space of the \textit{C.elegans} neural connectomes, which we refer to as joint graph inference. 
We focus on two aspects of joint graph inference: seeded graph matching and joint vertex classification.
\subsection{Seeded Graph Matching}
\begin{figure}
\centering
\includegraphics[width=0.5\textwidth]{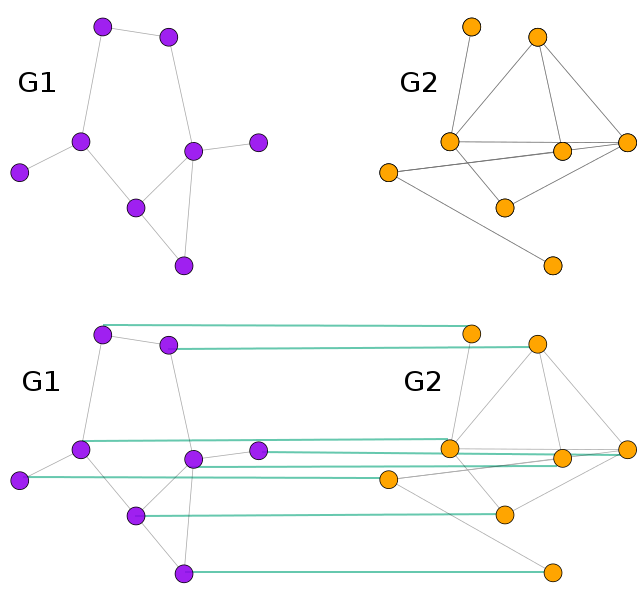}
\caption{\textbf{A depiction of graph matching.} Given two graphs $G_1$ and $G_2$, graph matching seeks an alignment (represented in the green lines) between the vertices across two graphs.}
\label{fig:gm_flow}
\end{figure}

The problem of seeded graph matching (SGM) is a subproblem of the graph matching (GM) problem, which has wide applications in object recognition \citep{berg2005shape, caelli2004eigenspace}, image analysis \citep{conte2003graph}, computer vision \citep{cho2012progressive, zhou2012factorized}, and neuroscience \citep{haris1999model, vogelstein2011fast, zaslavskiy2009global}. 
Given two graphs, $G_1 = (V_1, E_1)$ and $G_2 = (V_2, E_2)$ with respective adjacency matrices $A_1$ and $A_2$, and $|V_1| = |V_2| = n$, the GM problem seeks a bijection $\phi$ between the vertex sets that minimizes edge disagreements \citep{lyzinski2014divide, fishkind2012seeded}. 
The graph matching problem is NP-hard \citep{van1990handbook}. It is not known whether any graph matching algorithm is efficient, and it is suspected that none exist. 
For a comprehensive survey on the graph matching problem, see \cite{conte2004thirty} and \cite{vogelstein2011fast}. The intuitive idea of graph matching is seen in Figure \ref{fig:gm_flow}.

The seeded graph matching (SGM) problem employs additional constraint, where a partial correspondence between the vertices is known a priori. Those vertices are called ``seeds". Addition of seeds makes the graph matching problem has only a slight change in the graph matching algorithm, and improves the graph matching performance \citep{fishkind2012seeded}. 
Let $S_1 \subset V_1$ and $S_2 \subset V_2$ be two subsets of the vertex sets, and suppose $S_1 = S_2 = \{1,2,\dots, m\}$. The elements of $S_1$ and $S_2$ are called seeds, and the remaining $n-m$ vertices are non-seeds. In the SGM problem, one seeks a a bijection, with constraint on $S_1$ such that $\phi_{S_1} = \phi_{S_2}$, to minimize the number of induced edge disagreements. 

The SGM problem, as a subproblem of GM, is NP-hard. We seek an approximated SGM solution that is computationally efficient. The performance of SGM solutions is measured by the matching accuracy $\delta(m)$, defined as the number of correctly matched non-seeded vertices divided by the total number of non-seeds $n-m$. When the number of seeds $m$ is given, the remaining $n-m$ vertices need to be matched. Hence, the chance matching accuracy is $\frac{1}{n-m}$, and this accuracy increases as $m$ increases. For larger values of $m$, more information on the partial correspondence between the vertices is available, and thus the SGM matching accuracy becomes higher. In this work, we apply the state-of-the-art SGM algorithm developed by \cite{fishkind2012seeded}, seek the correspondence between the two types of neuron connectomes, and study the joint structure of the worm neural connectomes.

\subsection{Joint Vertex Classification}

When we observe the adjacency matrix $A \in \{0, 1\}^{n \times n}$ on $n$ vertices and the class labels $\{Y_i \}_{i=1}^{n-1}$ associated with the first $(n-1)$ training vertices, the task of vertex classification is to predict the label $Y$ of the test vertex $v$. In this case study, the class labels are the neuron types: motor neurons, interneurons and sensory neurons. 
In this work, we assume the correspondence between the vertex sets across the two graphs is known. Given two graphs $G_1 = (V,E_1)$ and $G_2 = (V, E_2)$ where $V = \{v_1, \dots, v_{n-1}, v\}$, and given the class labels $\{Y_i \}_{i=1}^{n-1}$ associated with the first $(n-1)$ training vertices, the task of joint vertex classification predicts the label of a test vertex $v$ using information jointly from $G_1$ and $G_2$. 

Fusion inference merges information on multiple disparate data sources in order to obtain more accurate inference than using only single source. Our joint vertex classification consists of two main steps: first, a fusion information technique, namely the omnibus embedding methodology by \cite{priebe2013manifold}; and secondly, the inferential task of vertex classification. 

\begin{figure}[h]
\centering
\includegraphics[width=0.66\textwidth]{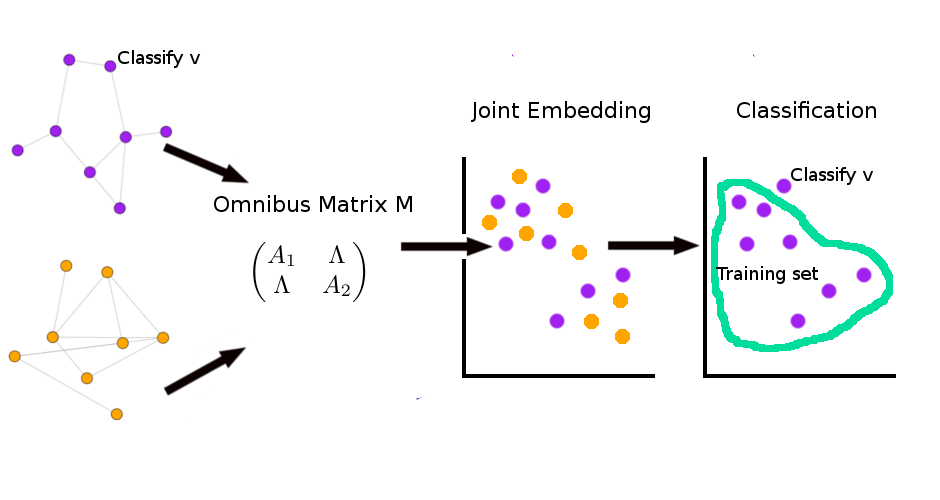}
\caption{\textbf{A depiction of joint vertex classification.} An illustration of joint vertex classification, which embeds the joint adjacency matrix -- the omnibus matrix, and classifies on the embedded space.}
\label{fig:jvc_flow}
\end{figure}

The step of omnibus embedding proceeds as follows. Given $G_1$ and $G_2$, 
we construct an omnibus matrix $M = \left( \begin{matrix}
A_1 & \Lambda \\
\Lambda & A_2  \end{matrix} \right) \in \mathbb{R}^{2n \times 2n}$, where $A_1$ and $A_2$ are the adjacency matrices of $G_1$ and $G_2$ respectively, and the off-diagonal block is $\Lambda = \frac{1}{2}(A_1 + A_2)$. We consider adjacency spectral embedding \citep{sussman2012consistent} 
of $M$ as $2n$ points into $\mathbb{R}^d$. Let $U = \left[\begin{matrix}
U_1 \\
U_2
\end{matrix} \right] \in \mathbb{R}^{2n\times d}$ denote the resulted joint embedding, where $U_1 \in \mathbb{R}^{n \times d}$ is the joint embedding corresponding to $G_1$, and $U_2 \in \mathbb{R}^{n\times d}$ to $G_2$. 
Our inference task is vertex classification. Let $\mathcal{T}_{n-1}: = U_1([n-1],:) \in \mathbb{R}^{(n-1) \times d}$ denote the training set containing the first $n-1$ vertices. We train a classifier on $\mathcal{T}_{n-1}$, and classify the test vertex $v$. A depiction of the joint vertex classification procedure is seen in Figure \ref{fig:jvc_flow}.

\begin{figure}[b]
\centering
\includegraphics[width=0.6\textwidth]{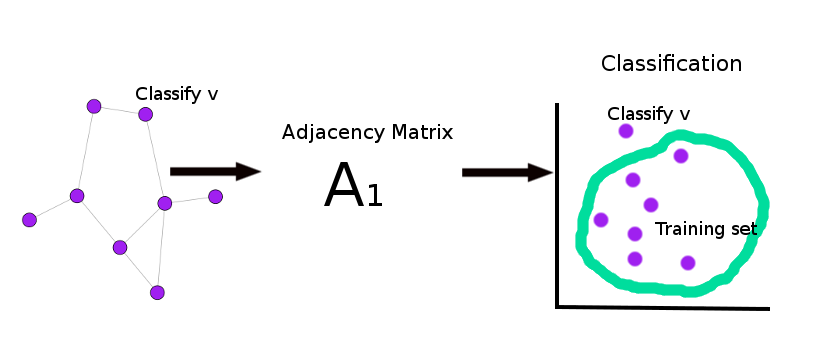}
\caption{\textbf{A depiction of single vertex classification.} An illustration of single vertex classification, which embeds one single adjacency matrix, and classifies on the embedded space.}
\label{fig:svc_flow}
\end{figure}

We demonstrate that fusing both pairs of the neural connectomes generates more accurate inference results than using a single source of connectome alone. We consider single vertex classification for comparison, which embeds the adjacency matrix $A_1$ to $\mathbb{R}^d$ via adjacency spectral embedding, and classifies on the embedded space. A depiction of the single vertex classification procedure is seen in Figure \ref{fig:svc_flow}.

\section{Discoveries from the Joint Space of the Neural Connectomes} \label{gm}

\subsection{Finding the Correspondence between the Chemical and the Electrical Connectomes}

\begin{figure}[h]
\centering
\includegraphics[width=0.5\textwidth]{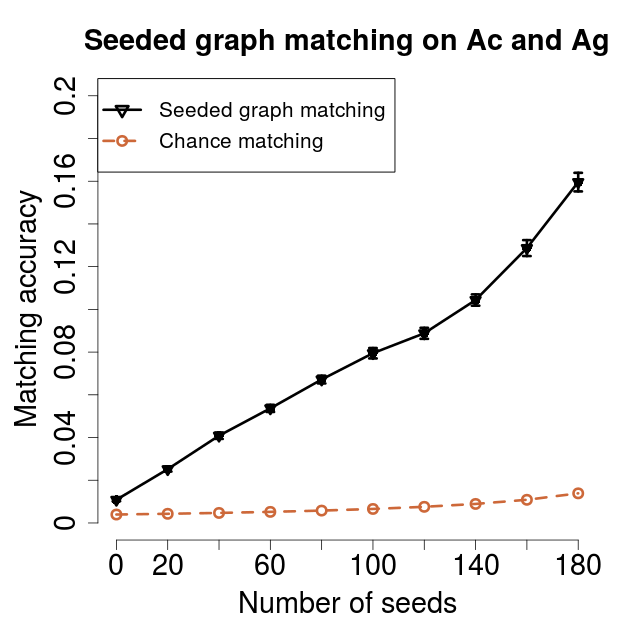}

\caption{\textbf{Seeded graph matching on the \textit{C.elegans} neural connectomes.} For each selected number of seeds $m \in \{0, 20, 40, \dots, 180\}$, we randomly select $100$ independent seeding sets $S_1$ and apply SGM for each Monte Carlo replicate. The SGM mean accuracy $\delta(m)$, plotted in black, is obtained by averaging the accuracies over the $100$ Monte Carlo replicates. As the number of seeds $m$ increases, the accuracy increases. The chance accuracy, plotted in brown dashed line, 
is much lower than the SGM accuracy. This suggests that a significant similarity exists between the two types of synapse connections. The SGM performance on the \textit{C.elegans} neural connectome is much more significant than chance but less than a perfect matching, indicating the optimal inference must proceed in the joint space of both neural connectomes.} \label{fig:sgm}
\end{figure}

We apply seeded graph matching on the paired \textit{C.elegans} neural connectomes, and discover the underlying structure preserved across the chemical and the electrical connectomes \citep{fishkind2012seeded}. Figure \ref{fig:sgm} presents the errorbar plot of the seeded graph matching accuracy $\delta(m)$, plotted in black, against the number of seeds $m\in \{0, 20, 40, \dots, 180\}$. For each selected number of seeds $m$, we randomly and independently select 100 seeding sets $S_1$. For each seeding set $S_1$ at a given number of seeds $m$, we apply the state-of-the-art seeded graph matching algorithm \citep{fishkind2012seeded}. The mean accuracy $\delta(m)$ is obtained by averaging the accuracies over the $100$ Monte Carlo replicates at each $m$. As $m$ increases, the matching accuracy improves. This is expected, because more seeds give more information, making the SGM problem less difficult. The chance accuracy, plotted in brown dashed line, at each $m$ is $\frac{1}{n-m}$, which does not increase significantly as $m$ increases. 

We note two significant neurological implications based on our graph matching result. First, SGM on the pair of connectomes indicate that the chemical and the electrical connectomes have statistically significant similar connectivity structure. The second significant implication is: If the performance of SGM on the chemical and the electrical connections were perfect, then one could consider just one (either one) of the paired neural connectomes without losing much information. If performance of SGM on the chemical and the electrical connections were no better than random vertex alignment, then it suggests that there is no structure similarity across the two connectomes, and this further suggests that analysis on the connectomes should proceed separately and individually. In fact, the seeded graph matching result on the \textit{C.elegans} neural connectome is much more significant than chance but less than a perfect matching. This demonstrates that the optimal inference should be performed in the joint space of the chemical and the electrical connectomes. This discovery is noted in \citep{lyzinski2014seeded}.

\subsection{Predicting Neuron Types from the Joint Space of the Chemical and the Electrical Connectomes}
The result of SGM on the \textit{C.elegans} neural connectomes demonstrates the advantage of inference in the joint space of the neural connectomes, and provides a statistical motivation to apply our proposed joint vertex classification approach. Furthermore, the neurological motivation of applying joint vertex classification stems from illustrating a methodological framework to understand the coexistence and significance of chemical and electrical synaptic connectomes. 

\begin{figure}
\centering
 \includegraphics[width=0.35\textwidth]{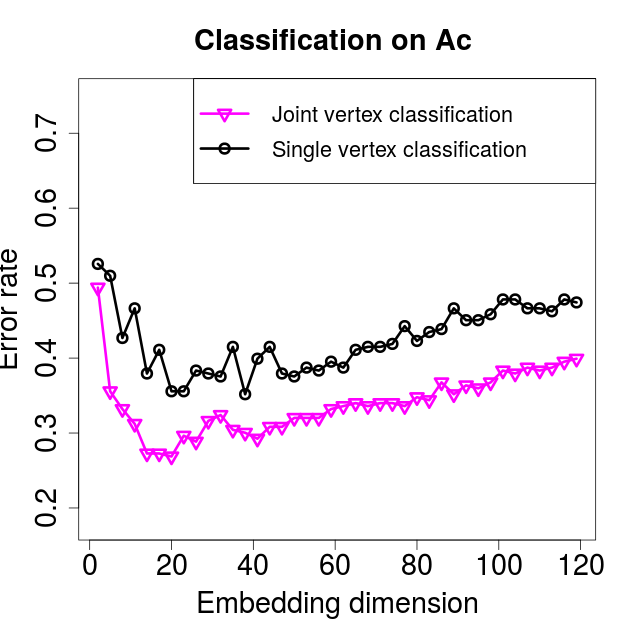}
\hspace{.4in}
 \includegraphics[width=0.35\textwidth]{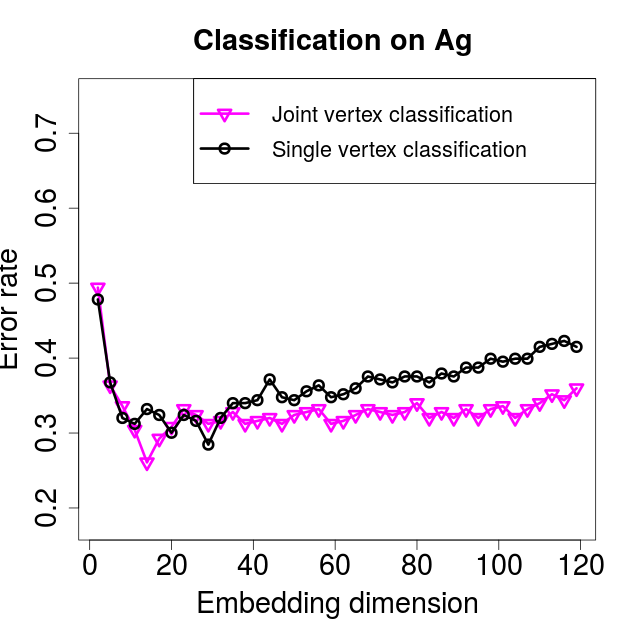}
\caption{\textbf{Classification performance of joint and single vertex classification.} (Left) Classification on the chemical connectome $A_c$. For all embedding dimensions $d \in \{2, 5, 8, \dots, 116, 119\}$, the error rate of joint vertex classification, plotted in magenta, is lower than the single vertex classification, plotted in black. (Right) Classification on the electrical gap junctional connectome $A_g$. For most of the embedding dimensions especially with larger values, the error rate of joint vertex classification, plotted in magenta, is lower than the single vertex classification, plotted in black.  
Our classification result indicates that using information from the joint space of the neural connectomes improves classification performance.}

\label{fig: jvc_svc_result}
\end{figure}

We apply joint vertex classification and single vertex classification on the paired \textit{C.elegans} neural connectomes, and compare the classification performance. The validation is done via leave-one-out cross validation. 
Here we do not investigate which embedding dimension $d$ or classifier are optimal for our classification task, and we choose support vector machine classifier with radial basis \citep{cortes1995support} for the classification step. 
The paired plots in Figure \ref{fig: jvc_svc_result} present the misclassification errors against the embedding dimensions $d \in \{2, 5, 8, \dots, 116, 119\}$. 
For the chemical connectome, the joint vertex classification (plotted in black) outperforms the single vertex classification (plotted in magenta) at all the considered embedding dimensions. For the electrical connectome, the joint vertex classification (plotted in black) outperforms the single vertex classification (plotted in magenta) at most of the considered embedding dimensions, especially larger valued embedding dimensions. 

The superior performance of the joint vertex classification over the single vertex classification has an important neuroscientific implication. In many animals, the chemical synapses co-exist with the electrical synapses. Modern understanding of coexistence of chemical and electrical synaptic connectomes suggest such a coexistence has physiological significance. 
We discover that using both chemical and electrical connectomes jointly generates better classification performance than using one connectome alone. This may serve as a first step towards providing a methodological and quantitative approach towards understanding the coexistent significance.

\section{Summary and Discussion}
The paired \textit{Caenorhabditis elegans} connectomes have become a fascinating dataset for motivating a better understanding of the nervous connectivity systems. We have presented the unique statistical approach of joint graph inference -- inference in the joint graph space -- to study the worm's connectomes. Utilizing jointly the chemical and the electrical connectomes, we discover statistically significant similarity preserved across the two synaptic connectome structures. Our result of seeded graph matching indicates that the optimal inference on the information-processing properties of the connectomes must proceed in the joint space of the paired graphs.

The development of seeded graph matching provides a strong statistical motivation for joint vertex classification, where we predict neuron types in the joint space of the paired connectomes. Joint vertex classification outperforms the single vertex classification against all embedding dimensions for our different choices of dissimilarity measures. Fusion inference using both the chemical and the electrical connectomes produces more accurate results than using one (either one) connectome alone, and enhances our understanding of the \textit{C.elegans} connectomes. 
The chemical and the electrical synapses are known to coexist in most organisms. Our proposed joint vertex classification provides a methodological and quantitative framework for understanding the significance of the coexistence of the chemical and the electrical synapses. Further development of joint graph inference is a topic of ongoing investigation in both neuroscience and statistics.
 
\section{Acknowledgement}
This work is partially supported by a National Security Science and Engineering Faculty Fellowship (NSSEFF), Johns Hopkins University Human Language Technology Center of Excellence (JHU HLT COE), the XDATA
program of the Defense Advanced Research Projects Agency (DARPA) administered through Air Force Research Laboratory contract FA8750-12-2-0303, and Acheson J. Duncan Fund for the Advancement of Research in Statistics. 

\bibliographystyle{plainnat}
\bibliography{Celegans.bib}
\end{document}